\documentclass{epl}

\title{Power-Law Size Distribution of Supercritical Random Trees}
\author{P. De Los Rios}
\institute{Institut de Physique Th\'eorique,
Universit\'e de Lausanne, \\ CH-1015 Dorigny-Lausanne, Switzerland.
}
\pacs{89.75.Da}{Systems obeying scaling laws}
\pacs{89.75.-k}{Complex systems}
\pacs{89.90.+n}{Other topics in areas of applied and interdisciplinary physics}

\begin{document}

\maketitle

\begin{abstract}
The probability distribution $P(k)$ of the sizes $k$ of
critical trees (branching ratio $m=1$)
is well known to show a power-law behavior $k^{-3/2}$. Such behavior corresponds
to the mean-field approximation for many critical and self-organized critical
phenomena. Here
we show numerically and analytically
that also supercritical trees (branching ration $m>1$)
are "critical" in that their size distribution obeys a power-law $k^{-2}$.
We mention some possible applications of these results.
\end{abstract}

\section{Introduction}
Ideal trees, either regular or random, play a great role in the description of
natural phenomena: many systems, from rivers~\cite{Rinaldo+,Rinaldo++} to blood vessels
and lungs~\cite{West+,Jay},
to real trees can be described by geometrical branched structures.
Other phenomena can be described by branched structures in time ({\it i.e.}
branching processes),
with cascades of events generating new events in a multiplicative fashion:
examples span from physics ({\it e.g.} nuclear chain reactions and directed percolation)
~\cite{Reggeon}, to
biology (speciation)~\cite{Burlando1,Burlando2} and many others disciplines.
More recently, trees (and branching processes) have been used to model the
mean-field approximation
of many non-equilibrium, self-organized critical systems such as the
Bak-Tang-Wiesenfeld sandpile and the Bak-Sneppen model~\cite{self1,self2}. The dynamics of
these models has a branching structure, where some local events can trigger
more events of the same kind, with some rules given by the branching probability
distribution ({\it i.e.} the probability $p_n$ to trigger $n$ new events)
and by the geometric constraints induced by the finite spatial dimension $d$
(that is, different branches can interact when coming to the same region).
To model branching processes in infinite dimension (mean-field) the usual
assumption is that there are no
interactions between different branches: the  statistical
properties of branching, $p_n$, are preserved, but any spatial constraints are
lost.
In particular, the mean-field limit of the above mentioned {\it critical}
models is recovered considering {\it critical}
branching processes~\cite{MF-BP1,MF-BP2}. A critical
branching process
is characterized by an average branching ratio $m = \sum_n n p_n = 1$,
so that every generation of
branching is (on the average) identical to the preceding ones.
What people are usually interested in
is the probability distribution $P(k)$ of the tree sizes $k$, that is, the
number of sites
(or of branching events, in a branching process jargon) making up the trees.
It is a well-known result of branching process theory that $P(k) \sim k^{-3/2}$
for critical trees~\cite{Harris}. This power-law behavior is consistent with the
assumption of criticality, as common wisdom suggests.

Interestingly enough, much less is known about the size distribution
of supercritical trees.
In this paper we show, through simulations and analytical arguments, that
also the supercritical case ($m > 1$) exhibits a power-law distribution of
tree sizes $P(k) \sim k^{-2}$. Some connections of this result to the structure
of the Internet and to taxonomic systems are proposed.

\section{Critical and Supercritical Trees: Numerical and Analytical Results}

\begin{figure}
\onefigure{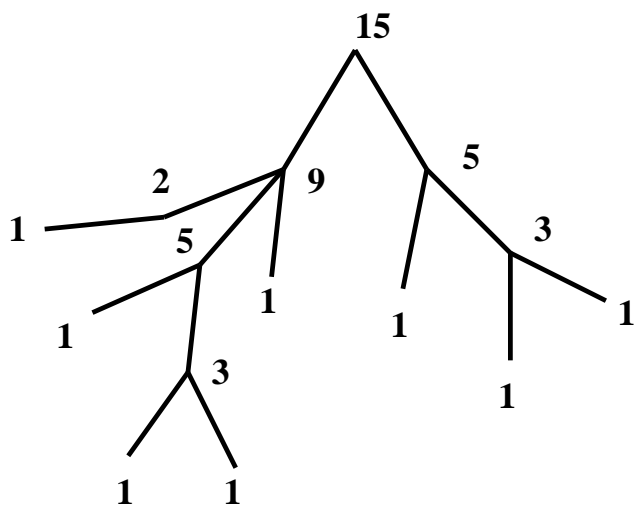}
\caption{A random Cayley tree grown for five generations (from $0$ to $4$).
The last generation sizes are set to $1$, and all the other subtree sizes are
also explicitly written.}
\label{fig1}
\end{figure}

Starting from a root site (generation $0$), a random tree is grown
letting every site at generation $t$ ($0 \le t < t_{max}$)
branch into $n$ new sites (generation $t+1$) with probability $p_n$.
An example of a random tree is shown in Fig.\ref{fig1}.
We are interested in the size distribution of the subtrees: picking
a site at random on the tree, what is the probability $P(k)$ that the subtree
that is rooted on it has size $k$? In Fig.\ref{fig1} such sizes are also
marked for every site. Sites at generation $t_{max}$ are assigned
a size $1$ (just themselves).

\begin{figure}
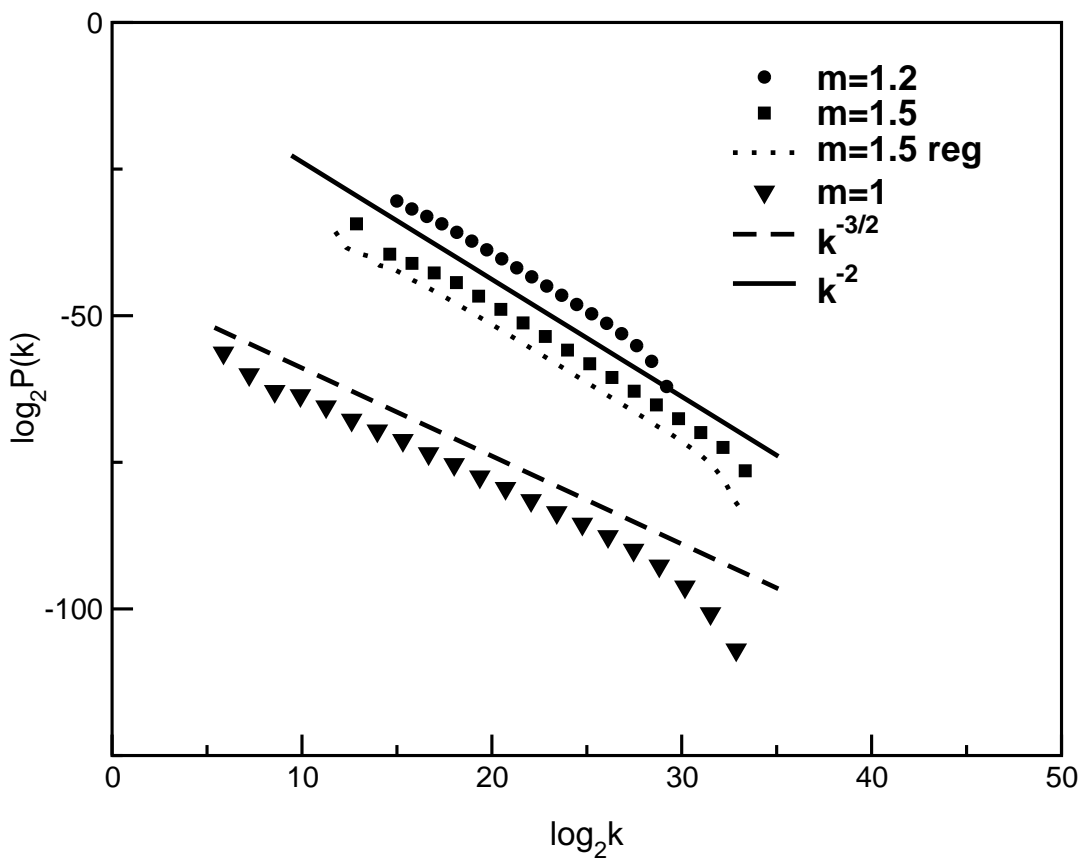

\onefigure{datibin.eps}
\caption{$P(k)$ for $m=1$, $m=1.2$ and $m=1.5$ (for this last case we
show the results both on random trees and on the regular exponentially
growing lattice: both exhibit the same behavior); the trees are grown for
$100$ ($m=1$), $50$ ($m=1.2$) and $30$ ($m=1.5$) generations, and averages
are taken over $10^5$ ($m=1,1.2$) and $10^3$ ($m=1.5$) realizations. The data
are binned on intervals growing as powers of $2$ 
(the $m=1.2$ data have also been shifted
to fit together into the same graph).}
\label{fig2}
\end{figure}

We have performed simulations for different
choices of $p_n$, both with $m=1$, finding the known result
$P(k) \sim k^{-3/2}$ and with $m > 1$,
finding $P(k) \sim k^{-2}$, as announced above (see Fig.\ref{fig2}).

To get an analytical insight in the origin of this behavior,
we must decompose $P(k)$ in generation dependent probabilities
$P_t(k)$ ($P_t(k)=0$ if $t>t_{max}$ since trees are grown
up to $t_{max}$).
We write explicitly the way in which generation dependent quantities
sum up to give the tree distribution $P(k)$:
\begin{equation}
P(k) = \left\langle \sum_{t=0}^{t_{max}} \frac{N_t(k)}{N_{tot}}
\right \rangle = \left\langle \sum_{t=0}^{t_{max}}
\frac{N_t}{N_{tot}} P_t(k) \right\rangle \;\;\;.
\label{distribution}
\end{equation}
In (\ref{distribution}) $N_t(k)$ is the number of sites at generation
$t$ that are roots of a subtree of size $k$, $N_t$ is the total number of sites
at generation $t$, $N_{tot}$ is the total number of sites on the tree.
$<\cdot>$ indicates the average over many realization of the system. Indeed,
$N_t$, $N_{tot}$ and $N_t(k)$ change from realization to realization.
As a first approximation we assume that, over many realizations,
the generation dependent quantities converge to their average values:
$N_t = m^t$,  $N_{tot} = (m^{t_{max}+1}-1)/(m-1)$.
Then we can write
\begin{equation}
P(k) = (m-1) \sum_{t=0}^{t_{max}} \frac{m^t}{m^{t_{max}+1}-1} P_t(k)
\simeq (m-1) \sum_{t=0}^{t_{max}} m^{-(t_{max}-t+1)} P_t(k)
\label{new distribution}
\end{equation}
where the last equality holds in the limit of large $t_{max}$.
We checked the reliability of such approximation,
simulating the same process on regularly
growing lattices. Starting at generation $0$ with $N_0$ sites, then
$N_t = m^t N_0$ (approximating it to the closest integer). Then each site
at generation $N_{t+1}$ is assigned to an ancestor in generation
$t$. The branching distribution is then the binomial
\begin{equation}
p_n(t) = \left(
\begin{array}{c}
N_{t+1} \\
n
\end{array}
\right)
\left(\frac{1}{N_t}\right)^n
\left(1-\frac{1}{N_t}\right)^{N_{t+1}-n}
\label{binomial}
\end{equation}
that in the limit $N_t \gg 1$ becomes the Poisson distribution
$p_n = e^{-m} m^n/n!$, independent from $t$.
This process gives the same results as the simulation of the
genuine random trees, that is $P(k)\sim k^{-3/2}$ if $m=1$,
$P(k) \sim k^{-2}$ if $m>1$, as shown in Fig.\ref{fig2}.

Problem (\ref{new distribution}) can be formally solved using
the recursion relation for the generating functions of $P_t(k)$.
Indeed we can write a relation between the probabilities $P_t(k)$ and
$P_{t+1}(k)$:
\begin{equation}
P_t(k) = \sum_n p_n \sum_{k_1=1}^k ...
\sum_{k_n=1}^k P_{t+1}(k_1)
\cdot ... \cdot P_{t+1}(k_n) \delta_{k_1+...+k_n+1,k}
\label{recursion in k}
\end{equation}

We define the generating functions $G_t(z) = \sum_k P_t(k) z^k$ and
$f(z) = \sum_n p_n z^n$. Then, multiplying both sides of (\ref{recursion in k})
by $z^k$ and summing over $k$, we obtain a recursion relation
for the generating functions at successive generations
\begin{equation}
G_t(z) = z f[G_{t+1}(z)]
\label{recursion in z}
\end{equation}
that is a very well known result from branching process theory~\cite{Harris}.
Eq. (\ref{recursion in z}) can be formally solved using the "initial condition"
$P_{t_{max}}(k) = \delta_{k,1}$ that translates into $G_{t_{max}}(z) = z$.
Then we can iterate Eq.\ref{recursion in z} back to generation $0$ to have an
explicit
expression of every $G_t(z)$, and,
from (\ref{new distribution})
obtain the generating function $G(z) = \sum_k P(k) z^k$. Knowing
$G(z)$ we could, in principle, get the asymptotic behavior of
$P(k)$: if $P(k) \sim k^{-\tau}$ for large $k$, then
$G(z) \sim 1 - c (1-z)^{\tau-1}$ for $z \to 1$. Unfortunately,
if $P(k) \sim k^{-2}$, then we could expect a non-analytic behavior
$G(z) \sim 1- c (1-z) \ln{(1-z)}$, that cannot be obtained
from an expansion of $G(z)$ for $z \to 1$. Instead, we should
expand around $z=0$, and resum the terms power-by-power of $z$,
which entails to computing explicitly $P(k)$ directly from
(\ref{new distribution}). Such an observation is of some relevance
since in the critical case $m=1$, invoking the time-translational invariance
of the process (that is, setting $P_t(k)=P_{t+1}(k)=P(k)$ so that
$G_t(z)=G_{t+1}(z)=G(z)$), Eq.\ref{recursion in z} becomes
$G(z) = z f[G(z)]$. Expanding both hands for $z \to 1$ and using the
corresponding form $G(z) \sim 1 - c (1-z)^{\tau -1}$, one can find
that to match all the powers of $(1-z)$ on both sides the only choice is
$\tau = 3/2$. Some simple cases for which $G(z) = z f[G(z)]$ can be
solved exactly are given by the branching probabilities
$p_n = 0$ if $n>2$, with $p_1+2 p_2=1$. In these cases the recursion equation
reduces to a second order equation in $G(z)$, that can be readily solved
to explicitely show that the leading non-analytic behavior for $z \to 1$
is given by $(1-z)^{1/2}$, hence $P(k) \sim k^{-3/2}$.

Still, even when the system is supercritical, the generating function formulation
is not fruitless. Indeed
it is possible to use it to compute the average subtree size
$<k> = \sum_k k P(k)$. Given the generating function $G(z)$
it is easy to see that the average size is
\begin{equation}
<k> = \left[ z \frac{d}{dz} G(z) \right]_{z=1} = (m-1) \sum_{t=0}^{t_{max}}
m^{-(t_{max}-t+1)} G_t'(1)\;\;\;.
\label{average value}
\end{equation}
Taking the derivative of (\ref{recursion in z}) we obtain a recursion relation
for $G'_t(1)$,
\begin{equation}
G'_t(1) = 1 + m G'_{t+1}(1)\;\;\;.
\label{mean recursion}
\end{equation}
With the "initial"
condition $G'_{t_{max}} = 1$,
we obtain
\begin{equation}
G'_t(1) = \frac{m^{t_{max}-t+1}-1}{m-1}
\label{recursive average}
\end{equation}
from which we finally get
\begin{equation}
<k> = \sum_{t=0}^{t_{max}} m^{-(t_{max}-t+1)} (m^{t_{max}-t+1}-1)
\sim t_{max}+1\;\;\;.
\label{average k}
\end{equation}

We find therefore that the average value $<k>$ diverges as $t_{max} \to \infty$,
a clear indication of a power-law behavior. Yet, it diverges as the
logarithm of the maximal allowed size, that is $k_{max} \sim m^{t_{max}+1}$,
which is the average total number of sites on the tree.
This is already a strong indication that asymptotically $P(k) \sim
k^{-2}$. 

\begin{figure}
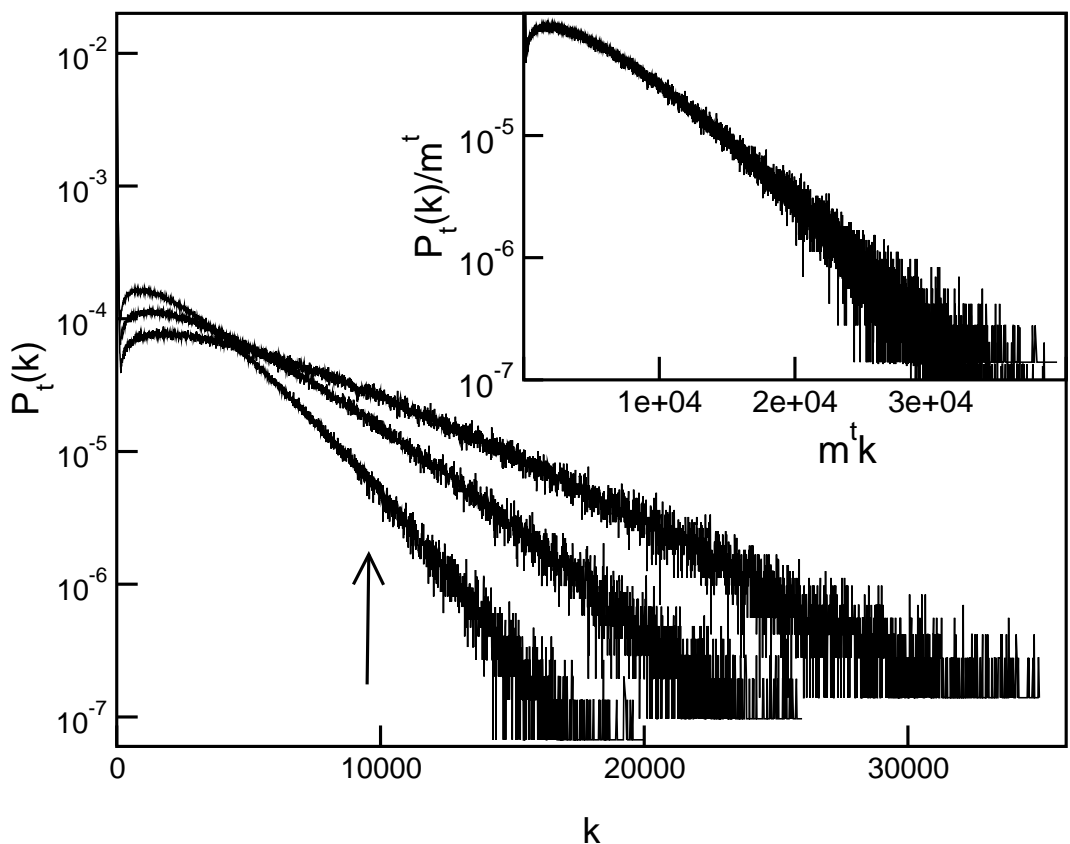

\onefigure{clas.eps}
\caption{Generation distributions $P_t(k)$ for $t=8,10,12$ from
upper to lower, respectively, in correspondence with the arrow.
The average branching ratio
is $m=1.2$, and trees are grown for $40$ generations. Averages are taken
over $10^6$ trees. In the inset the distributions rescaled according to
(10) are shown to nicely collapse.}
\label{fig3}
\end{figure}

We then look at the behavior of $P_t(k)$ from simulations.
We plot in Fig.\ref{fig3} the generation probabilities $P_t(k)$ at
generations $t=8,10,12$ (trees are grown for $40$ generations,
with $m=1.2$). They clearly decay exponentially for large values of $k$.
Such an exponential decay is suggestive of the presence of a size $k_t$
characteristic of each generation.
There is a simple and intuitive relation between $P_t(k)$ and
$P_{t+1}(k)$, namely,
\begin{equation}
P_{t+1}(k) = m P_t(m k)\;\;.
\label{scaling recursion}
\end{equation}
The relation between the arguments is readily understood thinking
that, on the average,
a site at generation $t$ is the root of a subtree of size $m k$ only
if each of its $m$ descendents at generation $t+1$ is the root of
a subtree of size $k$ (and formally it is expressed by Eq.\ref{mean recursion}).
The prefactor $m$ is due to normalization.
Interestingly, (\ref{scaling recursion}) is the generalization
to the supercritical case of the time-translational invariance
assumed to hold in the critical case $m=1$.
Relation (\ref{scaling recursion}) is numerically verified,
for $t \ll t_{max}$, where the asymptotic behavior has been reached,
as shown in the inset of Fig.\ref{fig3}.

Using (\ref{scaling recursion}) it is at last possible to have further evidence that
$P(k) \sim k^{-2}$ asymptotically.
We write
\begin{eqnarray}
P(m k) &=& (m-1) \sum_{t=0}^{t_{max}} m^{-(t_{max}-t+1)} P_t(m k) \nonumber \\
&=& (m-1) \sum_{t=0}^{t_{max}} m^{-(t_{max}-t+2)} P_{t+1}(k) \nonumber \\
&=& (m-1) \sum_{t=1}^{t_{max}} m^{-(t_{max}-t+3)} P_t(k) \simeq m^{-2} P(k)
\label{scalata}
\end{eqnarray}
where we have used $P_t(k)=0$ if $t>t_{max}$. The final equality is
consistent with a power-law decay with exponent $-2$ for large $k$.

To give further support to the idea that a functional form of $P_t(k)$
that satisfies (\ref{scaling recursion}) generically implies inverse square decay,
we assume the form
$P_t(k) = (1-a_t) a_t^{k-1}$ with $a_t = \exp({-m^{-(t_{max}-t)}})$,
that obeys (\ref{scaling recursion}).
Then
\begin{equation}
P(k) = (m-1) \left\lbrace  \sum_{t=0}^{t_{max}}
 a_t^{k-1} m^{-(t_{max}-t+1)} \right.
 -\left.\sum_{t=0}^{t_{max}}
 a_t^k m^{-(t_{max}-t+1)} \right\rbrace
\label{forma}
\end{equation}

The second sum on the {\it r.h.s} of (\ref{forma}) is
\begin{equation}
\sum_{t=0}^{t_{max}} a_t^k m^{-(t_{max}-t+1)} = \frac{1}{m}
\sum_{t=0}^{t_{max}} e^{-k m^{-(t_{max}-t)}} m^{-(t_{max}-t)} =
\frac{1}{m}\sum_{t=0}^{t_{max}} e^{-k m^{-t}} m^{-t}
\label{somma}
\end{equation}
Going from sums to integrals we write
\begin{equation}
\sum_{t=0}^{t_{max}}  e^{-k m^{-(t_{max}-t)}} m^{-(t_{max}-t)}
 \simeq 
\frac{1}{\ln{m}} \frac{1}{k} \int_{e^{-k}}^1 dy
\label{integrale}
\end{equation}
where we posed $y = e^{-k m^{-t}}$, such that, in the limit of large $t$, $\Delta y \sim m^{-t} \Delta t \to 0$
and the switch from sums to integrals is justified. 
Moreover, the upper limit of integration, $1$, comes after taking the
limit $t_{max} \to \infty$. After applying the same approximation
on the first sum  on the {\it r.h.s} of (\ref{forma}), eventually we have
\begin{equation}
P(k) \simeq \frac{1}{k-1} \int_{e^{-(k-1)}}^1 dy - \frac{1}{k} \int_{e^{-k}}^1
dy \sim \frac{1}{k^2}
\label{risomma}
\end{equation}
for large values of $k$.

A derivation of analogous results has been given in \cite{Noest} for trees with a fixed branching ratio.
Such a derivation based on the Mellin transform, although extremely
instructive, cannot be extended directly to variable branching ratios and in particular for
a non zero death probability.

\section{Conclusions}

We have studied the size distribution $P(k)$ of random trees
and subtrees. Alongside with the well known result $P(k) \sim k^{-3/2}$
for the case of critical trees (unitary branching ratio), we have found that
also supercritical trees show a power-law behavior $P(k) \sim k^{-2}$.
Some occurences of this behavior have already been found.

In taxonomy~\cite{Burlando1,Burlando2}, it has been found that the distribution of
taxa according to the number of their subtaxa has a power-law behavior, with exponents
scattered around $2$. Since taxonomy is intrinsically related to
a branch tree organization of genera, families, species etc., it is suggestive
to think that such exponents could be related to the tree-like nature
of the system itself, rather than to some underlying dynamical critical
process.

Even more recently it has been pointed out that such a power-law behavior is present also
for trees generated by connections on the Internet~\cite{GCalda}: there each node
of the tree is a site on the Internet, and the branches correspond to the sites it is linked
to (actually the structure of the Internet is that of a network, but search and trace programs
superimpose to it a corresponding tree structure). The number of sites on the Internet that
collect an area (number of sites) $k$ scales as $k^{-z}$ with $z$ close to $2$. Using this result we
can infer that the Internet, when looked at as a tree, is exponentially growing.

Recently, the inverse-square law has emerged, for similar reasons, 
in marine ecology~\cite{AmosPrivate}.

From a more general viewpoint, the emergence of "criticality" in such a simple and
well known framework as supercritical trees is suggestive of a whole
new family of systems (webs and supercritical webs) still hiding
interesting properties. Some preliminary simulations for directed webs show that
this seems to be the case.

\acknowledgments

This work has been partially supported
by the European Network contract FMRXCT980183 and by the Swiss National
Research Found FNRS 21-61397.00.

\end{document}